\pdfoutput=1
\documentclass[15pt]{article}

\usepackage{colortbl}
\definecolor{myblue}{rgb}{0.8,0.85,1}
\usepackage[preprint]{acl}

\usepackage{times}
\usepackage{latexsym}

\usepackage[T1]{fontenc}
\usepackage[utf8]{inputenc}

\usepackage{microtype}

\usepackage{inconsolata}

\usepackage{graphicx}

\usepackage{booktabs}  % 用于 \toprule, \midrule, \bottomrule
\usepackage{multirow}  % 用于 \multirow
\usepackage{array} 
\usepackage{hyperref}

\title{Bridging the Gap: \\Self-Optimized Fine-Tuning for LLM-based Recommender Systems}
\author{Heng Tang \\
  Zhejiang University \\
  Hangzhou, China \\
  \texttt{tangheng23@zju.edu.cn} \\
  \And
  Feng Liu \\
  OPPO Research Institute \\
  Shenzhen, China \\
  \texttt{liufeng4hit@gmail.com} \\
  \And
  Xinbo Chen \\
  \hspace{1cm}
  South China University of Technology \\
  Guangzhou, China \\
  \hspace{1cm}
  \texttt{202130140087@mail.scut.edu.cn}
  \AND
  Jiawei Chen \thanks{Corresponding author.} \\
  Zhejiang University \\
  Hangzhou, China \\
  \texttt{sleepyhunt@zju.edu.cn}
  \And
  Bohao Wang \\
  Zhejiang University \\
  Hangzhou, China \\
  \texttt{bohao.wang@zju.edu.cn}
  \And
  Changwang Zhang \\
  OPPO Research Institute \\
  Shenzhen, China \\
  \texttt{changwangzhang@foxmail.com}
  \AND
  Jun Wang \\
  OPPO Research Institute \\
  Shenzhen, China \\
  \texttt{junwang.lu@gmail.com}
  \And
  Yuegang Sun \\
  Intelligence Indeed \\
  Guangzhou, China \\
  \texttt{bulutuo@i-i.ai}
  \And
  Bingde Hu \\
  Zhejiang University \\
  Hangzhou, China \\
  \texttt{tonyhu@zju.edu.cn}
  \And
  Can Wang \\
  Zhejiang University \\
  Hangzhou, China \\
  \texttt{wcan@zju.edu.cn}
}
\begin{document}
\maketitle
\begin{abstract}
Recent years have witnessed extensive exploration of Large Language Models (LLMs) on the field of Recommender Systems (RS). There are currently two commonly used strategies to enable LLMs to have recommendation capabilities: 1) The "Guidance-Only" strategy uses in-context learning to exploit and amplify the inherent semantic understanding and item recommendation capabilities of LLMs; 2) The "Tuning-Only" strategy uses supervised fine-tuning (SFT) to fine-tune LLMs with the aim of fitting them to real recommendation data. However, neither of these strategies can effectively bridge the gap between the knowledge space of LLMs and recommendation, and their performance do not meet our expectations.

To better enable LLMs to learn recommendation knowledge, we combine the advantages of the above two strategies and proposed a novel "Guidance+Tuning" method called Self-Optimized Fine-Tuning (SOFT), which adopts the idea of curriculum learning. It first employs self-distillation to construct an auxiliary easy-to-learn but meaningful dataset from a fine-tuned LLM. Then it further utilizes a self-adaptive curriculum scheduler to enable LLMs to gradually learn from simpler data (self-distilled data) to more challenging data (real RS data). Extensive experiments demonstrate that SOFT significantly enhances the recommendation accuracy (37.59\% on average) of LLM-based methods. The code is available via \url{https://anonymous.4open.science/r/Self-Optimized-Fine-Tuning-264E}.
\end{abstract}

\begin{figure}
\centering
\includegraphics[width=0.5\textwidth]{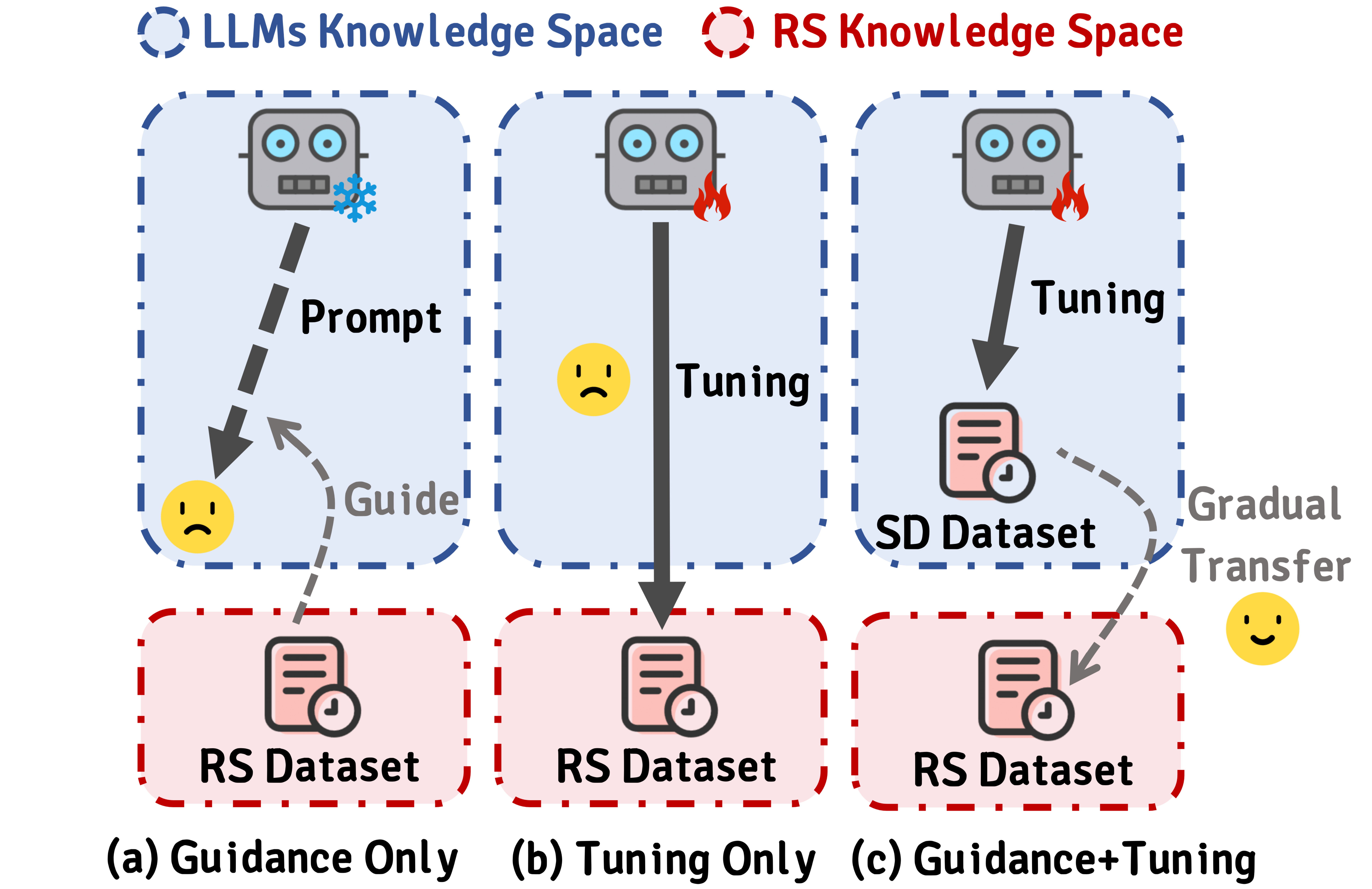}
\caption{Three different training strategies used in LLM-based recommender. The RS Dataset is the real recommendation dataset and the SD Dataset is dataset generated via self-distillation.}
\label{fig:my_label}
\end{figure}

\section{Introduction}
Given the knowledge-rich and semantic reasoning capabilities of Large Language Models (LLMs), directly applying them in Recommender Systems (RS) has garnered significant attention. There are two primary approaches to enable LLMs to have recommendation capabilities currently: (1) The "Guidance-Only" strategy (Figure 1a) aims to exploit and amplify the inherent abilities of LLMs to follow instructions and recommend items \cite{LLM4Rec-fs2} \cite{LLM4Rec-fs3}. Most existing works structure users’ historical interactions as language prompts and instructs LLMs to infer user preferences to predict future interactions. (2) The "Tuning Only" strategy (Figure 1b) enables LLMs to fit directly to the recommendation datasets \cite{bigrec} \cite{llara}. In this context, supervised fine-tuning (SFT) is commonly employed to supervise the LLMs to generate responses that align with the target items given the prompts of historical behaviors.

While both strategies have demonstrated certain strengths, they are far from meeting expectations as they do not always outperform traditional models \cite{allm}. This is because LLMs are not specifically pretrained for recommendation tasks, resulting in a significant gap between the knowledge space of LLMs and the recommendation, while neither of these strategies can effectively bridge this gap. The "Guidance-Only" strategy can only optimize within the inherent knowledge space of the LLMs. On the other hand, the "Tuning-Only" strategy faces optimization challenges as the LLMs need to learn both simple and complex recommendation knowledge from the datasets simultaneously.

Therefore, to effectively bridge the Knowledge gap between LLMs and recommendation, a more pragmatic approach is to use a "Guidance+Tuning" strategy (Figure 1c). This aligns with the concept of curriculum learning \cite{curriculum}, where the model first assimilates simpler knowledge—recommendation knowledge within its own knowledge space—prior to grappling with more complex external recommendation knowledge. This strategy can facilitate the optimization of LLMs, providing a better starting point in the basins of attraction for the descent procedure in parameter optimization. Stemming from this concept, we propose a novel approach, Self-Optimized Fine-Tuning (SOFT), which incorporates two key components:

\begin{figure}
\centering
\hspace*{-0.3cm}\includegraphics[width=0.5\textwidth]{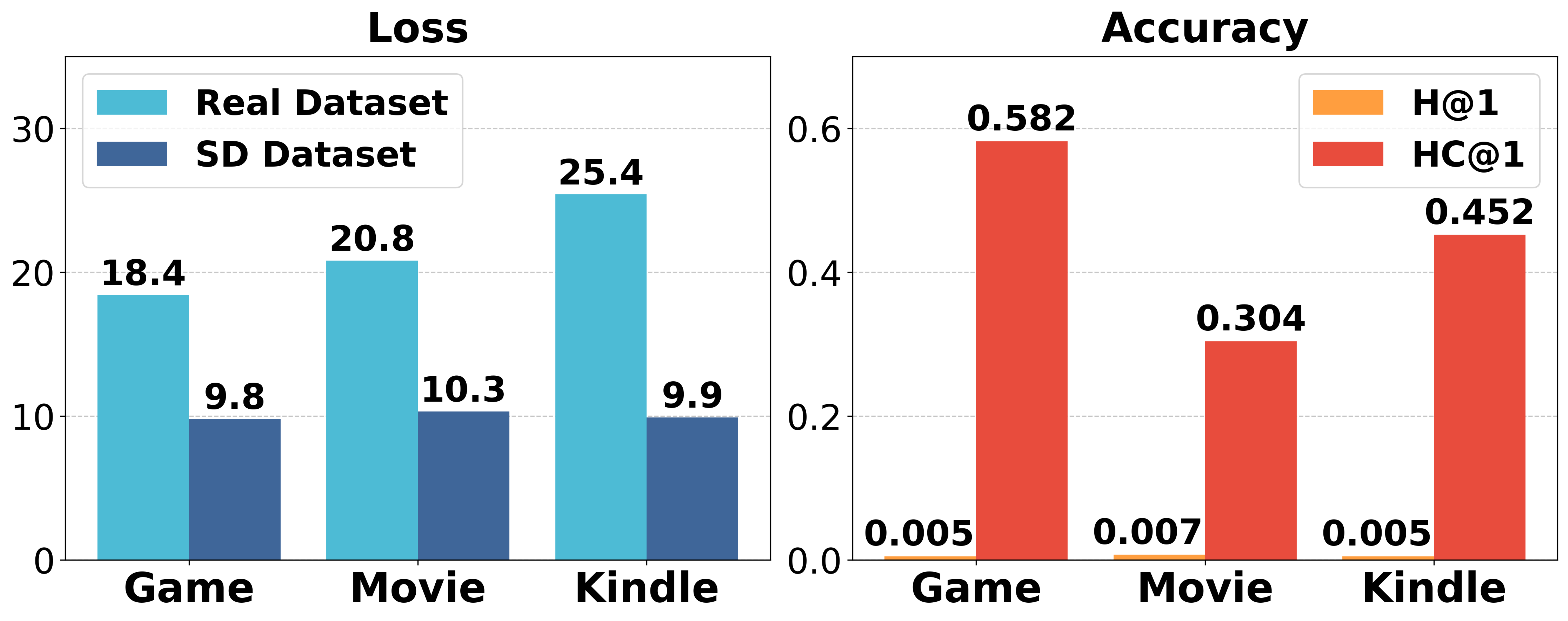}
\caption{(left) Comparison of the training loss on the real dataset and the SD dataset in the first epoch; (right) The accuracy of LLMs on the training dataset after SFT, where "H" refers to "Hit Ratio" and "HC" refers to "Hit Ratio of Category".}
\label{fig:my_label}
\end{figure}

\textbf{Auxiliary Data Generation via Self-Distillation for Guidance.} We use the data generated by the LLMs after SFT as auxiliary data to re-guide LLMs training. This process is akin to a thoughtful student reproducing their own written content, representing self-critique and further exploration of inherent abilities. We also conduct an empirical analysis to investigate the properties of this dataset. Firstly, as the dataset is generated by the LLMs itself, the knowledge it contains is more readily absorbed by the LLMs. As illustrated in Figure 2a, when LLMs are trained on a dataset generated via Self-Distillation (SD), the training loss in the initial epoch is only 47\% of that on the real RS dataset. Secondly, as illustrated in Figure 2b, even though the prediction accuracy of LLMs is quite low, with HR@1 being less than 1\% across three datasets, we were pleasantly surprised to find that the prediction accuracy of LLMs on item categories is quite substantial, averaging at 44.6\%. This suggests that, after SFT, LLMs have managed to learn some rudimentary recommendation knowledge, i.e., to analyze a user's historical interactions to discern item category preferences and recommend items from that category, though they still struggle to differentiate which item within that category would best suit the user. These two characteristics of the SD dataset make it an ideal candidate for serving as the "simpler" data in curriculum learning, without the need to manually define what constitutes simplicity.

\textbf{Self-Adaptive Curriculum Scheduler for Tuning.} After LLMs have absorbed knowledge that aligns with their own knowledge space, there is a need to learn more complex knowledge by gradually shifting their focus from easier data (SD data) to more challenging data (real RS data). An adaptive weighting mechanism is introduced, which adjusts the training focus based on the semantic distance between the LLM’s generated outputs and the target items. A larger distance indicates that the LLMs fall short on learning the recommendation knowledge, suggesting a need for easier data, while a smaller distance implies readiness to handle more complex data. This adaptive curriculum enables LLMs to progressively assimilate hard-to-learn recommendation knowledge with the aid of auxiliary self-distilled data.

\begin{figure}
\centering
\includegraphics[width=0.5\textwidth]{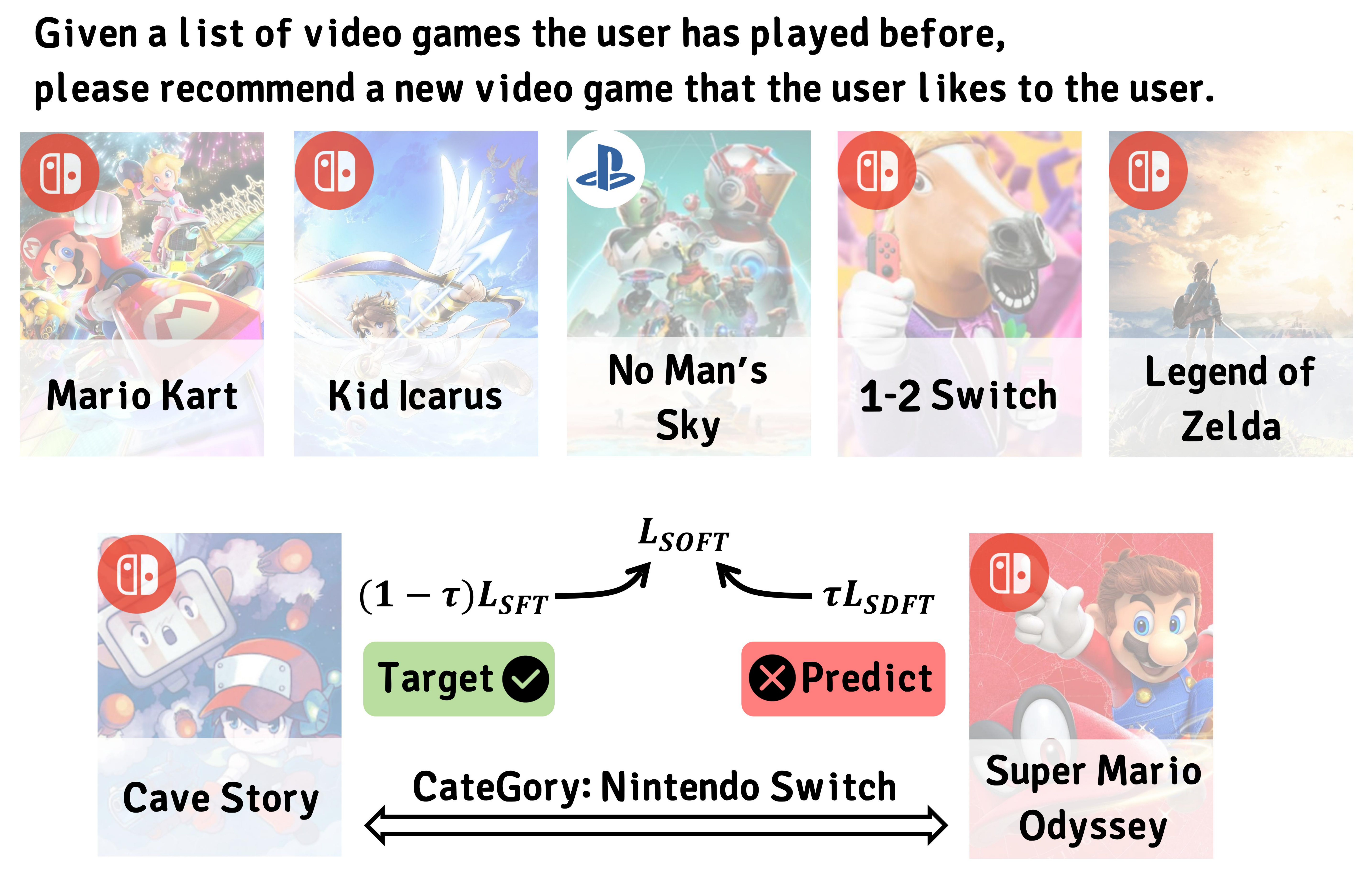}
\caption{An example of LLM's input and output. While the probability of the target item and the predict item being exactly the same is quite low, the probability of them being in the same category is relatively high.}
\label{fig:my_label}
\end{figure}

In summary, our contributions are as follows:
\begin{itemize}
    \item We identify the limitations of both the "Guidance Only" and "Tuning-Only" strategies and emphasize the importance of the "Guidance+Tuning" strategy to effectively bridge the gap between the knowledge space of LLMs and recommendation.

\item We propose a novel self-optimized fine-tuning strategy (SOFT), which leverages self-distillation to construct an auxiliary easy-to-learn dataset and employs self-adaptive scheduler to facilitate gradual learning from easy-to-difficult data.

\item We conduct extensive experiments to demonstrate that SOFT significantly improves the performance of LLM-based recommendation, achieving an average improvement of 37.59\%. 

\end{itemize}

\section{Preliminary}

\subsection{Task Formulation.} We define \(\mathcal{U}\) and \(\mathcal{I}\) as the sets of users and items, respectively. A user \(u\)'s historical sequence of interactions is denoted as \(S_u = [i_{u}^{1}, ..., i_{u}^{j}, ..., i_{u}^{L}]\), where each \(i_{u}^{j}\) is an item interacted with by user \(u\), listed in the order of occurrence. Here, \(u\) \(\in\) \(\mathcal{U}\), \(i_{u}^{j}\) \(\in\) \(\mathcal{I}\), and \(L\) represents the length of \(S_u\). The goal of the sequential recommendation task is to forecast the item \(i_{u}^{L+1}\) that will be preferred next by user \(u\), based on their historical interaction sequence.

\subsection{Supervised Fine-Tuning.} Supervised Fine-Tuning (SFT) stands as the predominant fine-tuning strategy employed in LLM-based recommenders. It leverages historical sequences as model inputs and designates the subsequent item as supervisory signals. Specifically, the training data, \(\mathcal{D}=\{(x_i,y_i)\}_{i=1,...,N}\), is constructed by integrating the textual information from the historical sequence \(S_u\) and an instruction prompt to form input \(x\). The content of the next item \(i_{u}^{L+1}\) is served as output \(y\). During the training process, the LLMs aim to increase the probability of each token in the label, given the tokenized prompt and the preceding tokens of the label, while simultaneously masking the tokens that follow. The corresponding loss function can be expressed as follows:

$$
L_{SFT} = -\sum_{(x_i, y_i) \in \mathcal{D} }\sum_{k=1}^{|y_i|} \log(\pi_\theta(y_{i,k} | x_i, y_{i,<k}))
$$

Despite the fact that SFT endows LLMs with the ability to make recommendations, they still grapple with the hallucination issue, manifesting as the generation of non-existent items.  \cite{bigrec} proposes a promising grounding strategy that identifies the item in the entire item space that has the closest semantic meaning to the one generated by the LLMs.  To elaborate, it defines \(z_{y}\) and \(z_{i}\) as the token embeddings of the generated descriptions and the descriptions of an item \(i\), respectively. It then computes the L2 distance used for grounding as follows:

$$
d_{yi} = ||z_{y} - z_{i}||^2
$$

\section{Methodology}

To effectively bridge the knowledge space gap between LLMs and recommendation, we combine the advantages of both "Guidance-Only" and "Tuning Only" strategies, and propose SOFT, a "Guidance+Tuning" approach. SOFT adopts the concept of curriculum learning, which imitates the progression found in human educational curricula. It first uses self-distillation to generate an easy-to-learn dataset to exploit and amplify the inherent recommendation abilities of LLMs. Subsequently, it employs a self-adaptive curriculum scheduler to progressively shift the training focus of the LLMs from easy to harder tasks.

\subsection{Auxiliary Data Generation via Self-Distillation}
Due to the substantial gap between the knowledge space of LLMs and recommendation, directly tuning LLMs using the real recommendation dataset can cause optimization difficulties, thereby weakening the final recommendation performance of the model. Therefore, it's necessary to tuning the LLMs with other data before using the real dataset. The suitable data needs to fulfill two requirements: (1) it should be \textbf{easy-to-learn}, not containing overly complex recommendation knowledge for LLMs; (2) it should be \textbf{meaningful}, containing some part of the recommendation knowledge found in the real dataset.

Traditional curriculum learning employs data-filtering along with a predefined difficulty measurer, artificially defining what data is simple for the model and filtering out such simple data from the the real dataset based on this definition. However, in recommendation tasks, defining simplicity in a comprehensive manner is challenging. Due to the powerful language understanding capabilities of LLMs, overly simplistic definitions can lead to overfitting of the LLMs to this subset of data, thereby impairing the performance of the model.

Hence, we resort to self-distillation to generate a simple dataset. More specifically, we employ LLMs after SFT to generate an output for each input, denoted as:

\[
\hat{y}_{i} = LLM_{SFT}(x_i)
\]

Subsequently, we construct a new auxiliary dataset \(\hat{\mathcal{D}}\), consisting of pairs of the form \((x,\hat{y})\). Compared to the real dataset \(\mathcal{D}\), this dataset more closely aligns with the distribution of the LLMs, making it easier for the model to learn. At the same time, although it is challenging for LLMs to learn all the knowledge in the recommendation dataset during the SFT process, they still manage to learn simple recommendation knowledge, such as making recommendations based on the categories of items from users' historical interactions. As shown in Figure 2, our empirical experiments also corroborate the above two points. This ensures that the dataset generated by self-distillation perfectly meets our requirements.

\begin{figure}
\centering
\hspace*{-0.3cm}\includegraphics[width=0.5\textwidth]{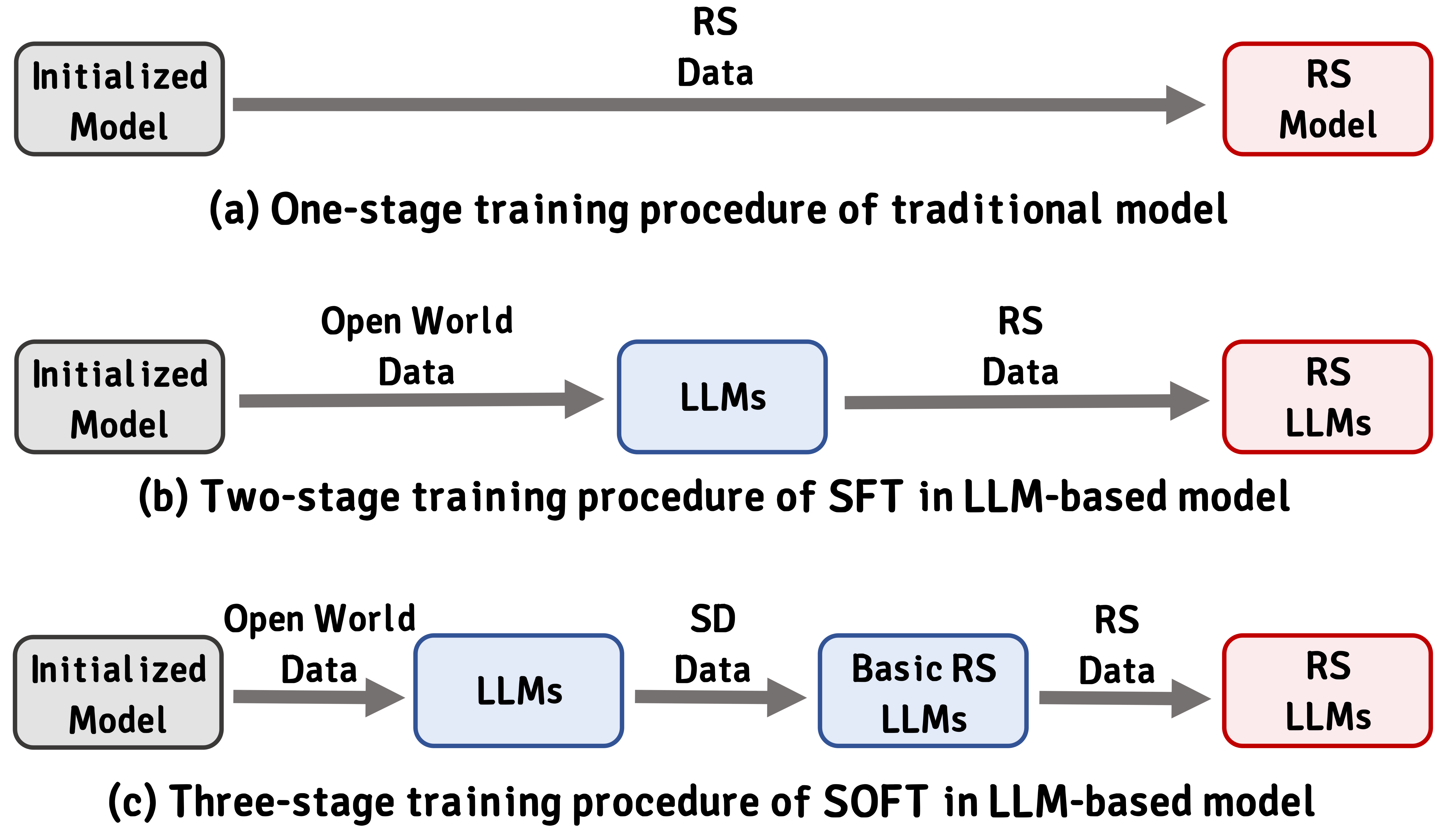}
\caption{The comparison of different training procedures among traditional models, SFT and SOFT.}
\label{fig:my_label}
\end{figure}

\subsection{Self-Adaptive Curriculum Scheduler}
Training solely on the SD dataset can only serve a guiding role. To effectively bridge the gap of knowledge space between LLMs and recommendation, it is necessary to introduce the real dataset for tuning, as it contains more complex and comprehensive recommendation knowledge.

A simple and effective method is to mimic human learning curriculums. While keeping the total amount of curriculum constant, we use a scheduler to control the ratio of difficulty in the curriculum. This allows the LLMs to gradually shift its attention from easier data to more challenging data during the fine-tuning process. As a result, we construct the SOFT loss function, which is formulated as follows:

$$
L_{SOFT} = (1-\tau)L_{SFT} + \tau L_{SDFT}
$$

$$
L_{SDFT} = -\sum_{(x_i, \hat{y_i}) \in \hat{\mathcal{D}} }\sum_{k=1}^{|\hat{y_i}|} \log(\pi_\theta(\hat{y}_{i,k} | x_i, \hat{y}_{i,<k}))
$$

This loss function incorporates a parameter \(\tau\) that acts as the curriculum scheduler to determine the weights of \(L_{SDFT}\) and \(L_{SFT}\). As training progresses, \(\tau\) gradually decreases, leading the LLMs to shift its focus from the easier SDFT loss to the harder SFT loss.

For the design of parameter \(\tau\), we do not adopt a predefined strategy due to its lack of flexibility across different tasks. Instead, we use the current learning situation of the LLMs to determine the difficulty of the curriculum, thereby designing a self-adaptive parameter \(\tau\) . Specifically, we utilize the average distance between the outputs of the current model and the real items to evaluate the learning state of the LLM. Before initiating each epoch of training, we first generate outputs \(y_t\) from the model and compute the average L2 distance of token embeddings between \(y_t\) and the real item \(y\):

$$
d_t = \sum^{M}_{i=1}d_{t,i} = ||z_{y_{t,i}}-z_{y_{i}}||^2
$$

We employ a sampling strategy to calculate the average distance, where \(M\) represents the number of samples in training data. This approach can significantly reduce the time required for model inference. This average distance serves as a measure of the deviation of the current model from the optimal one. As the training progresses and the distance decreases, we should assign more hard tasks to the LLMs and reduce the weight of \(\tau\).

We then employ a simple exponential function as the progression function, allowing \(\tau\) to be expressed as follows:

$$
\tau = e^{\alpha(\frac{d_t}{d_0}-1)}
$$

Where \(d_0\) represents the average distance at the commencement of the LLM's training and \(\alpha\) denotes a hyperparameter. It's worth mentioning that we have adopted the exponential function merely due to its simplicity. In practice, it can be replaced with other monotonically decreasing functions such as linear functions. In our experiments, we did not observe significant differences in performance. This is because, unlike the training of traditional models, the training of LLMs converges within a few epochs. With the \(\alpha\) being variable, the impact of using different progression functions on the shape of the \(\tau\) curve is relatively minor.

\subsection{Discussion: Curriculum Learning}
Curriculum learning \cite{curriculum} \cite{curriculumsurvey} mimics the process of human learning, in which the learning process is organized in a meaningful order. In curriculum learning, models first train on simpler (or smoother) data and then shift to more complex (or less smooth) data. This process has been demonstrated to be equivalent to unsupervised pre-training, which allows the training to to start in better basins of attraction of the descent procedure in parameter space, thereby achieving stronger generalization capabilities of the model.

As depicted in Figure 4, the two-stage training process (pretraining + SFT) of LLM-based recommenders can be interpreted as a specific form of curriculum learning. Initially, neural network models with randomly initialized parameters are pretrained using open-world data to acquire semantic understanding and reasoning capabilities, before being trained on more complex RS datasets. SOFT further decomposes SFT into a two-stage curriculum learning process, where LLMs are initially trained on the SD dataset to exploit and amplify their inherent basic recommendation capabilities, then subsequently trained on the real dataset to learn more complex recommendation knowledge.

\begin{table}[t]
\normalsize
\centering
\resizebox{0.5\textwidth}{!}{
\begin{tabular}{lcccc}
\toprule
 Datasets & \# Users & \# Items & \# Interactions & Density\\
\midrule
Game & 11658 & 5335 & 196508 & 0.32\%  \\
Movie & 67917 & 45578 & 1650120 & 0.05\% \\
Kindle & 76691 & 61528 & 2155790 & 0.05\% \\
\bottomrule
\end{tabular}
}
\caption{Statistic of the datasets.}
\label{tab:your_label}
\end{table}

\begin{table*}
\centering
\fontsize{7.5pt}{9pt}\selectfont
\setlength{\tabcolsep}{1pt}

\begin{tabular}{lccccccccccccccc}
\toprule
\multirow{2}{*}{Backbone} & \multirow{2}{*}{Method} & \multicolumn{4}{c}{Game} & \multicolumn{4}{c}{Movie} & \multicolumn{4}{c}{Kindle} \\
\cmidrule(lr){3-6}\cmidrule(lr){7-10}\cmidrule(lr){11-14}
 & & H@5 & H@20 & NG@5 & NG@20 & H@5 & H@20 & NG@5 & NG@20 & H@5 & H@20 & NG@5 & NG@20 \\
 \midrule
 & SASRec & 0.0150 & 0.0340 & 0.0074 & 0.0112 & 0.0120 & 0.0280 & 0.0082 & 0.0130 & 0.0100 & 0.0250 & 0.0058 & 0.0102 \\
 & DROS & 0.0160 & 0.0420 & 0.0091 & 0.0165 & 0.0190 & \textbf{0.0390} & 0.0101 & 0.0154 & 0.0140 & 0.0310 & 0.0076 & 0.0129 \\
\cmidrule(l){2-14}
 & LLM-CF & 0.0140 & 0.0380 & 0.0072 & 0.0114 & 0.0140 & 0.0300 & 0.0086 & 0.0136 & 0.0090 & 0.0280 & 0.0049 & 0.0087 \\
 & DLLM2Rec & 0.0150 & 0.0280 & 0.0097 & 0.0136 & 0.0100 & 0.0320 & 0.0046 & 0.0108 & 0.0080 & 0.0180 & 0.0045 & 0.0072 \\
\midrule
\multirow{6}{*}{BIGRec} & SFT & 0.0130 & 0.0310 & 0.0078 & 0.0127 & 0.0160 & 0.0200 & 0.0123 & 0.0134 & 0.0260 & 0.0320 & 0.0171 & 0.0188 \\
 & SDPO & 0.0150 & 0.0310 & 0.0090 & 0.0135 & 0.0100 & 0.0150 & 0.0079 & 0.0094 & 0.0180 & 0.0250 & 0.0110 & 0.0130 \\
 & SOFT(w/o SA) & 0.0170 & 0.0380 & 0.0107 & 0.0163 & 0.0220 & 0.0250 & \textbf{0.0197} & \textbf{0.0205} & 0.0260 & 0.0380 & 0.0181 & 0.0217 \\
 & SOFT & 0.0230 & 0.0470 & 0.0135 & 0.0200 & \textbf{0.0220} & 0.0250 & 0.0192 & 0.0200 & 0.0360 & 0.0460 & 0.0231 & 0.0261 \\
 \cmidrule(l){2-14}
 & \textit{Gain} & +53.33\% & +51.61\% & +50.00\% & +48.15\% & +37.50\% & +25.00\% & +56.10\% & +49.25\% & +38.46\% & +43.75\% & +35.09\% & +38.83\% \\
\midrule
\multirow{6}{*}{LLaRA} & SFT & 0.0110 & 0.0360 & 0.0080 & 0.0152 & 0.0170 & 0.0200 & 0.0135 & 0.0143 & 0.0340 & 0.0400 & 0.0214 & 0.0231 \\
 & SDPO & 0.0120 & 0.0360 & 0.0084 & 0.0153 & 0.0170 & 0.0200 & 0.0135 & 0.0143 & 0.0330 & 0.0390 & 0.0204 & 0.0220 \\
 & SOFT(w/o SA) & 0.0220 & 0.0490 & \textbf{0.0156} & 0.0228 & 0.0170 & 0.0210 & 0.0136 & 0.0147 & 0.0330 & 0.0440 & 0.0214 & 0.0246 \\
 & SOFT & \textbf{0.0230} & \textbf{0.0510} & 0.0150 & \textbf{0.0230} & 0.0200 & 0.0210 & 0.0160 & 0.0163 & \textbf{0.0380} & \textbf{0.0460} & \textbf{0.0245} & \textbf{0.0269} \\
  \cmidrule(l){2-14} 
 & \textit{Gain} & +91.67\% & +41.67\% & +78.57\% & +50.33\% & +17.65\% & +5.00\% & +18.52\% & +13.99\% & +11.76\% & +15.00\% & +14.49\% & +16.45\% \\
\bottomrule
\end{tabular}
\caption{Performance comparisons of SOFT with existing fine-tuning strategies on the two LLM-based recommender backbones, as well as with existing LLM-enhanced and traditional recommenders. The best performance is highlighted in bold. The term \textit{Gain} denotes the improvement of SOFT over the other fine-tuning strategies on the two backbones.}
\label{tab:your_label}
\end{table*}

\section{Experiments}
\label{sec: Benchmark}

In this section, we verify the effectiveness of SOFT by answering the research questions as follows:
\begin{itemize}
    \item\textbf{RQ1} : How does SOFT perform compared with LLM-based and  traditional recommender models?
    \item\textbf{RQ2} : What is the impact of self-adaptive curriculum scheduler on the performance of SOFT?
    \item \textbf{RQ3}: What is the impact of the different hyperparameter \(\alpha\) on the performance of SOFT?
    \item \textbf{RQ4}: How does SOFT compare with other methods in terms of training time?
\end{itemize}

\subsection{Experiment Settings}
\noindent\textbf{Datasets.} We conduct our experiments on three subsets of the Amazon reviews dataset: Video Games, Movies and TV, and Kindle Store\footnote{\url{https://amazon-reviews-2023.github.io/}}. These datasets encompass both user behavior sequences and item names, providing a comprehensive view of user-item interactions. To ensure the quality of our datasets, we perform a filtering step where users and items with fewer than 10 interactions are removed. Next, we arrange the interaction sequences in ascending order of timestamps and partition each dataset into training, validation, and testing sets, following an 8:1:1 ratio. For all datasets, we sample 1,000 sequences for testing and 512 sequences for validing. For all of the LLM-based methods, considering of the time consumption, we sample 4096 sequences for training. The dataset statistics are presented in Table 1.

\noindent\textbf{Baselines.} Our implementation of SOFT is conducted on two commonly used LLM-based recommender backbones: \textbf{BIGRec} \cite{bigrec} is a LLM-based recommender that formulates the recommendation task using natural language prompts and employs instruction-tuning for fine-tuning LLMs. \textbf{LLaRA} \cite{llara} leverages a novel hybrid prompting method that integrates ID-based item embeddings, learned by traditional recommendation models, with textual item features. 

We compare SOFT with the following categories of baselines:
\begin{itemize}
\item\textbf{Fine-tuning Strategy:} \textbf{SFT} is the most widely used fine-tuning strategy, which directly uses real items as supervisory signals to fine-tune LLMs. \textbf{SDPO} \cite{sdpo} is a strategy that further incorperates direct preference optimization (DPO) \cite{dpo} building on the basis of SFT.

\item\textbf{LLM-Enhanced Recommender:} \textbf{LLM-CF} \cite{LLMCF} utilizes LLMs to produce a foundation of thought chains, which are subsequently harnessed to augment sequential recommendation models. \textbf{DLLM2Rec} \cite{dllm2rec} is a method that distills knowledge from large, cumbersome LLM-based recommendation models to lightweight conventional sequential models, addressing the challenge of high inference latency in LLMs.

\item\textbf{Traditional Recommender:} We utilize the representative traditional recommender models, \textbf{SASRec} \cite{sasrec} and \textbf{DROS} \cite{dros}, in our experiments.
\end{itemize}
\noindent\textbf{Evaluation Metrics.} In our experiments, we adopt two widely-used metrics to assess the performance of our model: Hit Ratio (H) and Normalized Discounted Cumulative Gain (NG). Both metrics are calculated at cut-off values of 5 and 20 (K=5, K=20). We evaluate the model's performance using the all-ranking protocol, where all items that a user has not interacted with are considered potential candidates.

\noindent\textbf{Implementation Details.} For all LLM-based methods, we select Llama3.2-3B \cite{llama3} as the backbone with the learning rate set to 1e-4 and fine-tune LLMs using LoRA \cite{lora} on 4 Nvidia A800 GPUs. We run up to 7 epochs for both SFT and SOFT, and an additional 3 epochs for SDPO, with early stopping patience to 2. Our batch sizes are 256, 256, 128 for Game, Kindle, Movie datasets, respectively. For SOFT, we set the samples for computing average distance \(M\) to 256 and adjust the hyperparameter \(\alpha\) within the range of \{0.1, 1, 10, 100\}. For LLaRA, LLM-CF, and DLLM2Rec, we use SASRec as the backbone of their traditional models. For traditional models, we follow \cite{llara}, employing Adam optimizer, with a batch size of 256, an embedding dimension of 64, a learning rate of 0.001 and a grid search in [1e-3, 1e-4, 1e-5, 1e-6, 1e-7] for the coeffcient of L2 regularization. We conduct five experiments and took the average of the final results.

\begin{figure}
\centering
\hspace*{-0.3cm}\includegraphics[width=0.5\textwidth]{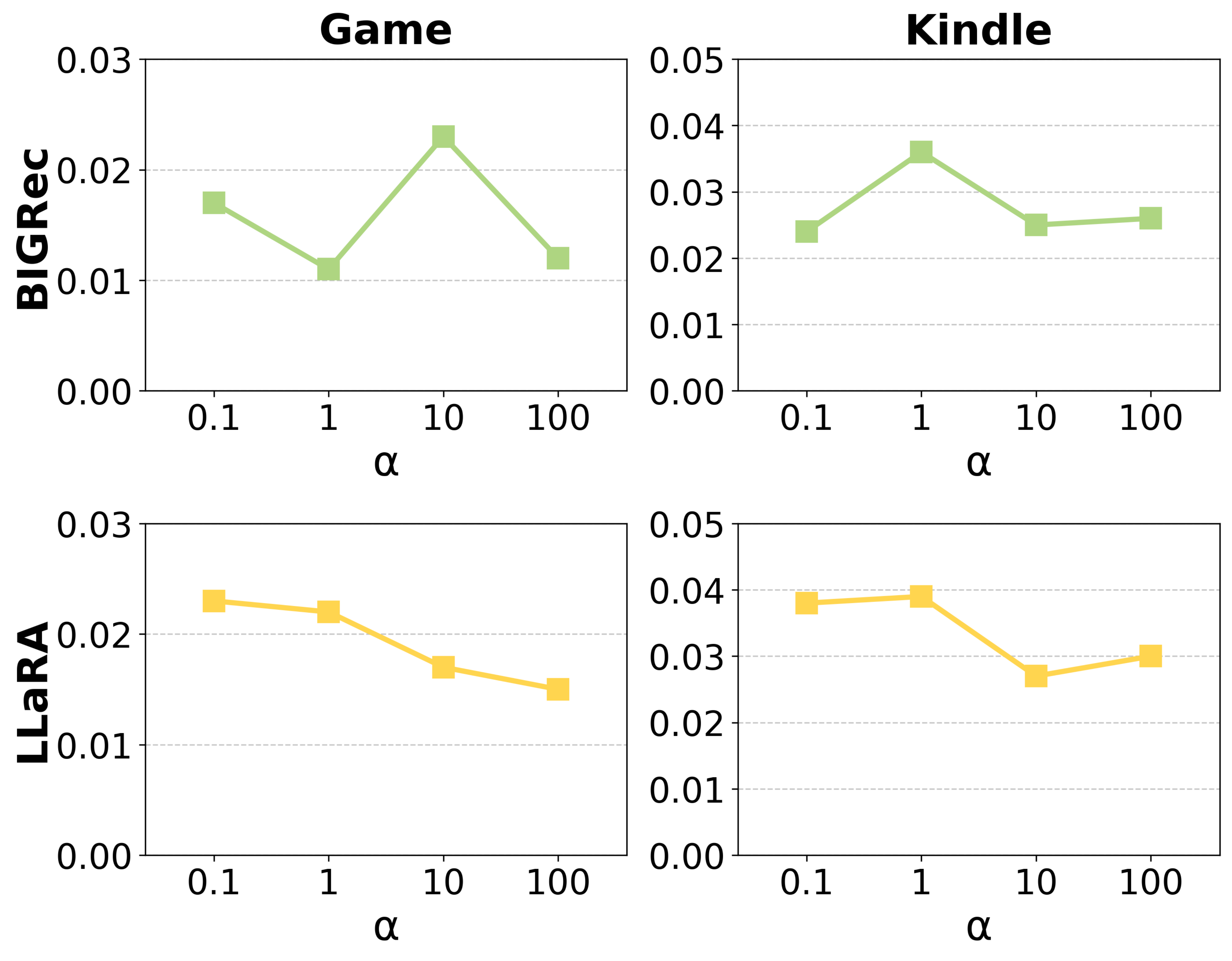}
\caption{The performance of SOFT with difference hyperparameter \(\alpha\). }
\label{fig:my_label}
\end{figure}

\subsection{Performance Comparison (RQ1)}
We compare our method with other LLM fine-tuning strategies, LLM-based recommendation models, and traditional recommendation models to demonstrate the effectiveness of our approach. The results of these comparisons are summarized in Table 4.

\textbf{Comparing with existing LLM fine-tuning strategies.} Regardless of whether BIGRec or LLaRA is used as the LLMs backbone, SOFT can outperform other fine-tuning strategies on most of the datasets and metrics, demonstrating the effectiveness of our method. We also observe that SDPO, a strategy that further fine-tunes on the basis of SFT, does not perform as well as SFT on some datasets. This could be because, in our all-ranking setting, negative samples include not only other items but also invalid tokens. Strategies like SDPO, which select negative samples through random sampling, may lead to a likelihood decrease phenomenon \cite{nca}, causing the probabilities of both positive and negative samples to drop simultaneously.

\textbf{Comparing with existing LLM-based recommendation models.} Our SOFT method, when using BIGRec as the backbone, outperforms LLaRA, LLM-CF, and DLLM2Rec, thereby demonstrating the effectiveness of our approach. When compared to LLaRA, which employs SFT for optimization, our method of applying SOFT at the loss level offers superior performance over the approach of adding ID-based item embeddings at the input end. In the case of LLM-CF and DLLM2Rec, where the LLMs merely serve a supporting role, their performance remains constrained by that of the base model, SASRec.

\textbf{Comparing with existing traditional recommendation models.} The optimal performance of SOFT surpasses that of the best-performing traditional recommendation model, DROS, across all datasets, demonstrating the effectiveness of our method. It's noteworthy that traditional models use more training data than LLMs. This suggests that LLM-based recommenders have a lower dependency on the volume of data, and when optimized with the appropriate loss, they can exhibit a higher performance ceiling than traditional recommendation models.

\begin{figure}
\centering
\includegraphics[width=0.4\textwidth]{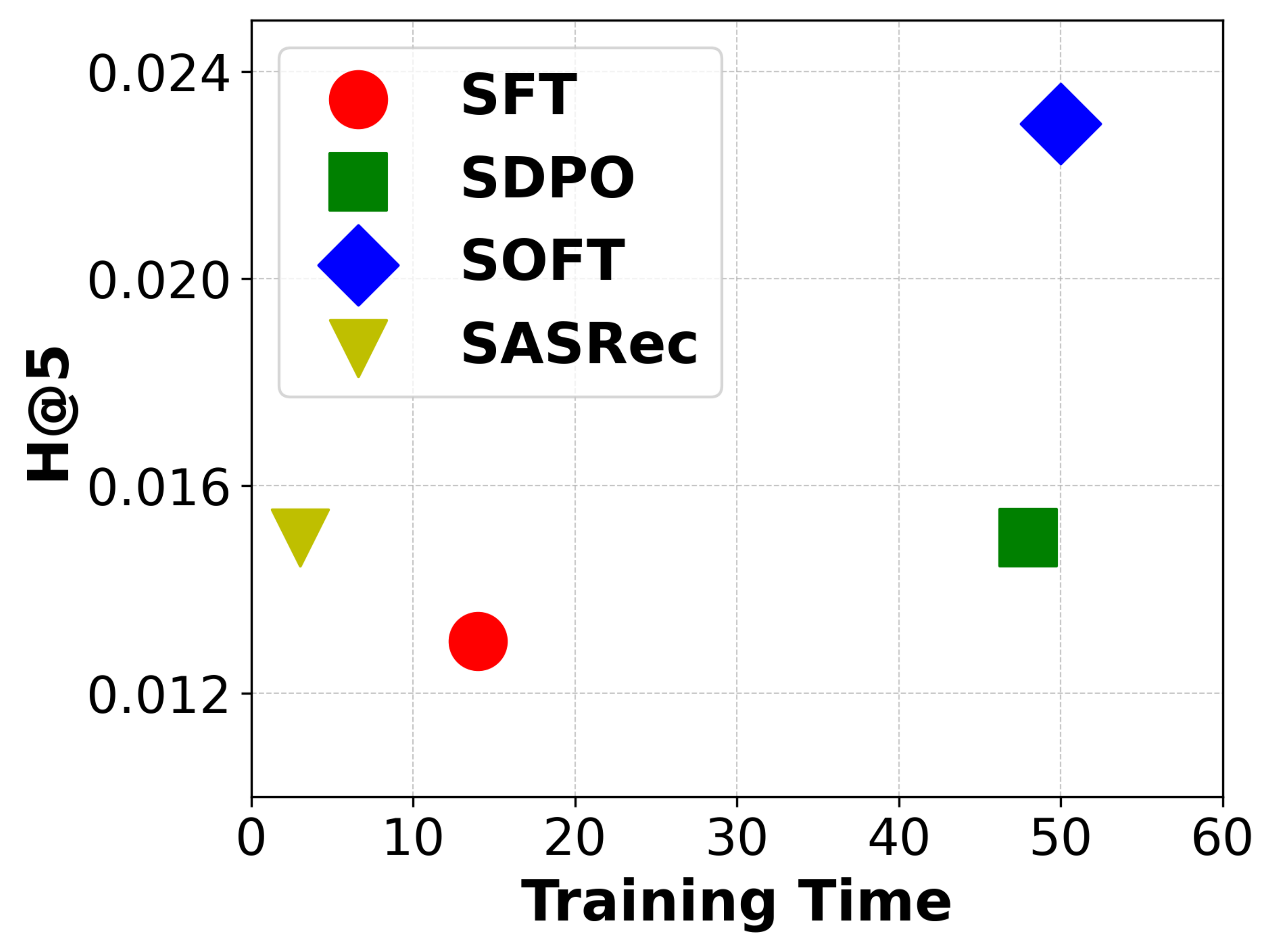}
\caption{The training time (minutes) and performance of SASRec and three LLMs fine-tuning strategies.}
\label{fig:my_label}
\end{figure}

\subsection{Ablation Study (RQ2)} To understand the impact of the self-adaptive curriculum scheduler module on SOFT, we conduct an ablation study on different datasets. Specifically, we set \(\tau\) to be constantly 1, allowing the LLMs to train solely on the SD dataset (w/o SA). This reduces the SOFT approach from a "Guidance+Tuning" method to a "Guidance-only" method. The results are presented in Table 2.

We found that training only on the SD dataset yields improvements across multiple datasets compared to SFT trained solely on the real dataset. This indicates that the outputs of LLMs after SFT can filter out more complex knowledge in the real dataset. Training with the SD dataset allows LLMs to focus more on learning simple knowledge suitable for them, thus guiding the LLMs to exploit and amplify their inherent recommendation capabilities. However, training only on the SD dataset does not always outperform SFT, reflecting the own advantages of both "Guidance-Only" and "Tuning-only" strategies. Compared to training on individual datasets, SOFT shows significant improvements, indicating that it can integrate the advantages of both "Guidance-Only" and "Tuning-only" strategies, thereby better bridge the gap of the knowledge space between LLMs and recommendation.

\subsection{Impact of hyperparameter (RQ3)} In this section, we conduct experiments to investigate the impact of different hyperparameter \(\alpha\)  selections on the performance of SOFT. We perform these experiments on two datasets, using both BIGRec and LLaRA as backbones. We evaluate the results using the H@5 metric, and the results can be found in Figure 5.

We find that selecting an appropriate alpha has a noticeable impact on the performance improvement of SOFT, indicating that setting an appropriate course curve can lead to significant performance enhancements in LLM fine-tuning. Simultaneously, we also observe that the optimal hyperparameters differ significantly across various datasets and backbones, underscoring the necessity of hyperparameter search.

\subsection{Analysis of Training Time (RQ4)} In this section, we evaluate and compare the training time of SOFT with other methods. All experiments are conducted on the Game dataset using a single Nvidia A800 GPU, with BIGRec searving as the backbone for LLM-based methods. The results of these comparisons can be found in Figure 6.

We observe that LLM-based methods significantly increase the training time compared to the traditional model, SASRec. However, using SFT alone does not result in a significant performance improvement compared to SASRec. Although our SOFT method further increases the training time compared to SFT, its training time is essentially on par with other complex fine-tuning strategies such as SDPO. Nevertheless, the performance of SOFT far exceeds that of all other methods, indicating that the increase in training time indeed yields substantial benefits.

\section{Related Work}
\subsection{Sequential Recommendation}
Sequential recommendation predicts the users' next item of interest based on their historical interaction. In recent year, deep learning models have been widely used in sequential recommendation. For instance, GRU4Rec \cite{GRU4Rec} uses RNN, Caser \cite{caser} uses CNN, SURGE \cite{SURGE} uses GNN, while SASRec \cite{sasrec} and BERT4Rec \cite{bert4rec} uses attention mechanism. And DROS \cite{dros} leverage distributional robust optimization in sequential recommendation and achieves state-of-the-art performance.
\subsection{LLM-based Recommendation System}
LLMs can be directly used for recommendation. "Guidance-Only" strategy directly employs pre-trained LLMs as the backbone for recommendation, aiming to explore their zero-shot capabilities \cite{LLM4Rec-fs1} \cite{LLM4Rec-fs2} \cite{LLM4Rec-fs3} \cite{LLM4Rec-fs4}, while "Tuning-only" strategy uses recommendation data to fine-tune LLMs in order to achieve better performance \cite{allm} \cite{LLM4Rec-ft1} \cite{LLM4Rec-ft2} \cite{LLM4Rec-ft3} \cite{LLM4Rec-ft4}. Additionally, some studies do not directly use LLMs for recommendations. Instead, they leverage the capabilities of LLMs to enhance traditional recommendation models, such as converting user and item semantic information into embeddings or using them as additional knowledge bases \cite{LLM4Rec-enhance1} \cite{LLM4Rec-enhance2} \cite{LLM4Rec-enhance3} \cite{LLM4Rec-enhance4}.
\subsection{Self-Distillation in LLM Fine-Tuning}
Self-Distillation (SD) is a process in which the model serves as both the teacher and the student, enabling the model to learn from itself \cite{sd}. As the capabilities of LLMs have improved, SD has started to be used in the field of Natural Language Processing (NLP) for LLM fine-tuning. Some studies have used SD and a small amount of external supervision to enable LLMs to generate their own instructions and responses as training data \cite{self-instruct} \cite{self-align}. Furthermore, some studies have leveraged the inherent summarization and reflection capabilities of LLMs. \cite{cot1} \cite{cot2} uses Chain-of-Thought (CoT) to allow LLMs to reflect on their own incorrect responses, \cite{SDFT} designs prompts to enable LLMs to generate responses that better fit the model's distribution. Our SOFT method specifically addresses the unique issue in recommendation systems where directly learning recommendation tasks is overly hard for LLMs. We use SD to enable LLMs to generate an easy-to-learn dataset for guidance, and thus improving the performance of LLMs in recommendations.

\section{Conclusion}
In this paper, we highlight the limitations of the most commonly used strategies in LLM-based recommendation systems: "Guidance-Only" and "Tuning-Only". Neither of these strategies can effectively bridge the gap between the knowledge space of  LLMs and recommendation systems, resulting in performance that falls short of expectations. To combine the respective strengths of these two strategies, we propose a "Guidance+Tuning" approach, Self-Optimized Fine-Tuning (SOFT), which incorporates the principles of curriculum learning. SOFT initially employs self-distillation to generate an easy-to-learn but meaningful dataset to guide the LLMs to exploit and amplify their inherent basic recommendation capabilities. Subsequently, it uses a self-adaptive scheduler to adjust the training focus of LLMs from simple to challenging data based on their current learning state. extensive experiments demonstrate that SOFT can significantly improve the performance of LLM-based recommendation.

\section*{Limitations}
Despite our SOFT model demonstrating promising performance across multiple datasets and backbone models, we must acknowledge the following limitations of our approach.

(1) Recommendation Task: Our method is solely focused on sequence recommendation tasks and only provides the names of items with which the user has previously interacted.

(2) Fine-Tuning Approach: As with most previous works, our method only uses LoRA for efficient parameter fine-tuning, and we have not adjusted the parameters.

Our future work will strive to address these limitations, further enhancing the performance and versatility of SOFT.

\bibliography{custom}

\appendix

\end{document}